\documentclass[12 pt]{article}
\usepackage[dvips]{graphicx}
\usepackage{amssymb}

\newcommand \nion {V}
\newcommand \qed{\vrule height5pt width5pt}

\newcommand \preprint {1}

\newcommand \no {\noindent}
\newcommand \sumxy {\sum_{<x,y>}}
\newcommand \hop {c^\dagger_x c_y }
\newcommand \nbad {N}

\newtheorem{lemma}{Lemma}
\newtheorem{theorem}{Theorem}

\begin{document}

\title{Periodic Ground States in the Neutral Falicov-Kimball Model in 
Two Dimensions}

\author{Karl Haller 
\\Department of Mathematics
\\University of Virginia
\\Charlottesville, VA 22903
\\ email:  kh8q@virginia.edu
\\
\\Tom Kennedy
\\Department of Mathematics
\\University of Arizona
\\Tucson, AZ 85721
\\ email: tgk@math.arizona.edu
\bigskip
}

\maketitle

\begin{abstract}
We consider the Falicov-Kimball model in two dimensions in the 
neutral case, i.e, the number of mobile electrons is equal to 
the number of ions. For rational densities between $1/3$ and $2/5$
we prove that the ground state is periodic if the strength of the 
attraction between the ions and electrons is large enough. 
The periodic ground state is given by taking the one dimensional 
periodic ground state found by Lemberger and then extending it 
into two dimensions in such a way that the configuration is 
constant along lines at a 45 degree angle to the lattice directions.
\end{abstract}


\bigskip

\vspace{\fill}
\hrule width2truein
\smallskip
{\baselineskip=12pt
\noindent
\copyright\ 2000 by the authors. Reproduction of
this article is permitted for non-commercial purposes.
\par}

\newpage

In the Falicov-Kimball model there are two types of particles. The 
electrons hop between nearest neighbor sites of the lattice while
the ions do not hop at all. The electrons do not interact with each
other except for the fact that they obey Fermi statistics. 
At most one ion is allowed at each lattice site and there is an 
on-site attraction between the ions and electrons. 
The Hamiltonian is 
\begin{equation} 
 H = \sumxy \hop - 4 U \sum_x c^\dagger_x c_x \nion_x \label{eqham}
\end{equation}
where $c^\dagger_x$ and $c_x$ are creation and annihilation
operators for the electrons. 
$\nion_x$ is the occupation number for the ions, i.e., $\nion_x= 1$ 
if there is a ion at $x$ and $\nion_x=0$ if there is not. 
The sum over $<x,y>$ is over nearest neighbor bonds in the lattice.
(The factor of 4 in front of the $U$ is included for latter convenience.)
The interaction of the electrons with the ions and the kinetic 
energy and fermi statistics of the electrons lead to an effective interaction
for the ions. We will study the ground states of 
this model on the square lattice in the neutral case
(the numbers of ions and electrons are equal) when the parameter $U$ 
is large and positive.
A review of rigorous work on the Falicov-Kimball model may be found 
in \cite{gm}.

In any number of dimensions
the ground state for density $1/2$ is the checkerboard configuration
for all $U>0$ \cite{bs,kl}. In two dimensions with large $U$ 
the ground states for densities 
$1/3, 1/4$ and $1/5$ are known rigorously and are periodic \cite{gjlii,k}.
Numerical results indicate periodic ground states for numerous other
values of the density \cite{watlem}.
In one dimension Lemberger \cite{lem}
showed that for any rational density, the ground
state is the periodic arrangement of the ions which is ``most homogeneous.''
(He gives an explicit algorithm for
determining the most homogeneous configuration.)
His proof applies when $U>U_0$ where $U_0$ depends on the 
denominator of the rational density.

The above results suggest that in two dimensions 
in the neutral case with large $U$ the 
ground state is always periodic. However, Kennedy \cite{ksep}
proved that for densities between $1/5$ and $1/4$ other than $2/9$
the ground state is not periodic.  
In this density range there is phase separation. 
The ground states for densites $1/5$, $2/9$ and $1/4$ are periodic, but
in between the ground state is formed by adjoining regions with 
these periodic configurations so as to produce the desired density.
Haller \cite{dissert} proved that the same phenomenon occurs for densities 
between $1/6$ and $2/11$.
So an analog of Lemberger's result cannot hold in two dimensions for 
all densities. In this paper we prove that an analog of his result
does hold for rational densities between $1/3$ and $2/5$. 
In this interval of densites the ground state is periodic.
It is constant along lines of slope $\pm 1$ (one choice for the 
entire ground state),  
and its restriction to any horizontal or vertical line is the same as the 
one dimensional ground state found by Lemberger. 
We expect that this result is true for the entire density interval
$1/3$ to $1/2$, but for technical reasons we cannot prove it for 
densities between $2/5$ and $1/2$.  

There is a physical argument that suggests that 
the neutral model with rational density wants to spread out the 
ions as much as possible.
Each electron spends most of its time at a site with an ion
since the attraction between electrons and ions is large. 
Now consider the effect of the kinetic energy of an electron. If nearby sites
have ions, then the electrons at those sites will restrict the movement 
of the electron we are considering. So its kinetic energy is 
minimized by arranging the ions so as to maximize the space that each electron 
has to move in. However, Watson \cite{watson} 
emphasized that there will typically 
be a mismatch between the lattice and the ion configuration that maximizes 
this space in the absence of a lattice.
Thus the lattice structure can frustrate the exclusion
principle's attempts to put the ions in the ``most homogeneous'' configuration.
In one dimension this frustration does not occur. In two dimensions
Kennedy's and Haller's results on phase separation show it does occur for 
some intervals of density, while this paper shows it does not for
other intervals. There does not appear to be any easy way to predict 
which intervals of density will have periodic ground states and 
which ones will have phase separation.

\ifodd \preprint 
{
\begin{figure}[density]
        \includegraphics{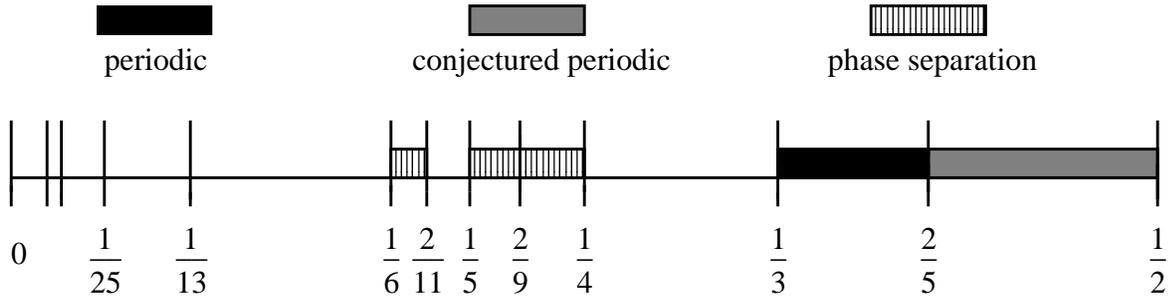}
        \caption{The known results for the two dimensional neutral model with 
                 large $U$ shown as a function of the density. 
                 The ground state is known to be periodic 
                 at all the densities at which there is a vertical line.  
                 In the interval $(0,1/6)$ the vertical lines indicate 
                 the first few densities $1/(n^2 + (n+1)^2)$ where the 
                 ground state is known to be periodic.  
		}
        \label{density}
\end{figure}
}
\else { } \fi

The results for the neutral model on the square lattice with 
large $U$ are summarized in figure \ref{density}.
The obvious open problem is to determine whether there is a periodic 
ground state or phase separation in these intervals where the 
behavior is unknown. However, the more important problem is to 
understand when there is phase separation vs. periodicity 
without the extensive calculations required in this paper or the 
papers on phase separation. 

The results on periodicity and phase separation in the square lattice
depend on extensive calculations of an effective Hamiltonian and 
thus depend heavily on the lattice structure. One possible way to 
gain some insight into the general periodic vs. phase separation 
question is to investigate other lattices. There 
are rigorous results on periodic ground states on
the triangular lattice for some densities \cite{triangular}.
In three dimensions, nothing is known with the  
exception of density $1/2$.

The preceding discussion and indeed all of this paper is only 
concerned with ground states. For density $1/2$ and any $U>0$ it is 
known that there are at least two Gibbs states if
the temperature is sufficiently low \cite{bs,kl} 
and the number of dimensions is at least two. 
For densities $1/3,1/4$ and $1/5$, a ``quantum'' 
Pirogov-Sinai theory must be used to prove that for 
large $U$ there are multiple Gibbs states at low temperature
\cite{frob,mess,mm,miracle}.  
The key ingredient in these approaches is a Peierls estimate. One 
can prove such an estimate for densities $1/6, 2/11$ and $2/9$ 
and so the methods should apply to these densities as well. 
For densities in $(1/3,2/5)$ we do not know how to prove the needed 
Peierls estimate, and so the existence of multiple Gibbs states 
at low temperatures is an open problem for these densities.

We now turn to our main result. The precise statement is as follows.

\smallskip

\begin{theorem} Suppose $\rho = p/q \in [1/3,2/5]$ where $p$ and $q$ are 
relatively prime. We assume the lattice $\Lambda$ is $L\times L$ where $L$ 
is a multiple of $4q$, and we impose periodic boundary conditions. 
Then there exists $U(q) >0$ such that for all
$U \ge U(q)$, the ground states are translations, reflections, and 
rotations of the configuration $S$ defined by the following properties: 
$S$ is constant along lines of slope $+1$, and the restriction of $S$ 
to any vertical or horizontal line is the same as the one dimensional 
large $U$ ground state 
with density $\rho$.
\label{pthrm}
\end{theorem}

First we give an overview of the proof.
When $U$ is large one may do perturbation theory in $1/U$. 
We start by considering $3\times 3$ blocks of sites and consider the 
perturbation series for the energy up to fourth order.  
Kennedy has shown that at this order, 
each block which looks like one of those shown in 
figure \ref{3blocks} contributes the same amount of energy, while any other
block contributes a higher amount of energy. So if a configuration exists
in which every $3\times 3$ block looks like one of those from figure
\ref{3blocks}, it minimizes $H$ through fourth order. It turns out that
such configurations do exist. In fact, Watson \cite{watson} has shown that 
they correspond to tilings by squares and parallelograms as shown in figure
\ref{vtypes}. In such a tiling the configuration must be constant on lines
of slope $\pm1$, where the sign for the entire configuration is determined
by the slope of the short sides of one of the parallelograms. 

Up to rotations and reflections, the ions in such a tiling must be one of
the three types shown in figure \ref{vtypes}. We may write the energy 
through eighth order in terms of the number of each vertex type. Through
sixth order we find that every tiling has the same energy. At eighth order,
tilings with only type B and C vertices have the same energy, but type
A vertices increase the energy. So if there exists configurations that 
correspond to square-parallelogram tilings with no type A vertices, then
they minimize $H$ through eighth order. Such configurations do
exist for densities in $[1/3,2/5]$.    

When we include the higher order terms, the most we can say at this point is
that the ground state consists primarily of large regions corresponding
to a square-parallelogram tiling with no type A vertices. Because of the
lack of type A vertices, these regions must contain parallelograms.
Furthermore, the slopes of the short sides of these parallelograms must
be the same throughout each individual region. This means that within each 
region, the configuration must be constant on lines of slope $+1$ or $-1$, 
but not both, as it would in a region consisting entirely of type A vertices.

\ifodd \preprint 
{
\begin{figure}[thp]
        \includegraphics{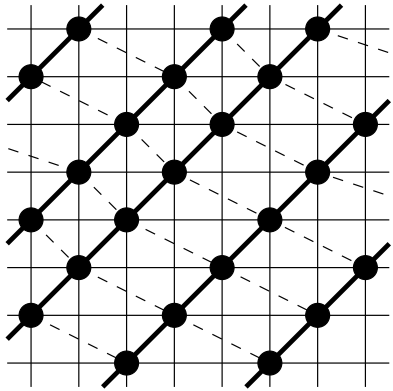}
        \caption{A typical example of a piece of one of the ``$+$'' regions.
		The thicker lines are the ``stripes'' of occupied sites.
		The dotted lines are inserted to show the squares and
		parallelograms.}
        \label{stripex}
\end{figure}
}
\else { } \fi

On each of these ``$+$'' and ``$-$'' regions, the problem
essentially becomes one-dimensional, since we are now only concerned with
the spacing between these ``stripes'' of occupied sites (see figure 
\ref{stripex}). It is tempting
to try to apply a procedure to each region, similar to that of Lemberger 
for the one-dimensional case, in order to determine the cheapest stripe
arrangement. However, there are two difficulties that arise. First there is
the problem of the shape of these regions. Their boundaries may be very 
irregular whereas an extension of Lemberger's method would require 
applying periodic boundary conditions to a fairly regular shape such as a
square or rectangle. Second there is 
a density constraint. The density inside the $+$ and $-$ regions need not 
be the same as the overall density. This is a problem because how
large $U$ needs to be in Lemberger's argument depends on the density.
Even if we could say something about the configuration 
in each $+$ and $-$ region, this would not take into account the ions 
lying outside these regions, so we effectively lose track of the original 
ion density. We must therefore find more regular regions to work with
that allow us to keep track of the original density.

The first step in overcoming these difficulties is to partition the lattice
into diamonds containing 8 sites, as shown in figure \ref{blocks}.
If one of these diamonds intersects a $3\times 3$ block not shown in
figure \ref{3blocks}, 
or it contains a type A ion, it may be associated with an 
increase in energy at eighth order. 
We call such diamonds ``bad.''  
The eighth order terms actually occur with $U^{-7}$, so the total energy
cost of the bad diamonds is at least $O(U^{-7}) \nbad$, 
where $\nbad$ is the number of bad diamonds. 
Each of the remaining ``good'' diamonds is contained in one of the $+$ or
$-$ regions mentioned above. In a $+$ region ($-$ region), we group the 
diamonds into $-45^{\circ}$ strips ($+45^{\circ}$ strips) in such a 
way that the strips are bounded on each of the short sides by a bad diamond. 
This is shown in figure \ref{rowcol}. 

If we were to apply periodic boundary conditions to a particular strip and 
compare the resulting energy with that of the 
energy of the strip in the original configuration, 
it would seem that this difference would be proportional to the
size of the boundary of the strip. However,
consider one of the $45^{\circ}$ strips of $-$ diamonds. Inside the strip,
the configuration is constant on lines of slope $-1$. 
These lines of constant configuration 
persist through the long sides of the strip, and continue in $S$ until they
encounter a bad diamond. So if there are no bad diamonds near the 
long sides, then imposing periodic boundary conditions on the strip
is effectively the same as the boundary conditions the strip had to 
begin with. This allows us to show that the total difference between
the energies of the strips in the original configuration
and their energies with periodic boundary conditions is 
$O(U^{-9}) \nbad$.

We then glue all the strips of good diamonds together with the bad diamonds
end to end to form one long thin strip. We apply periodic boundary conditions
to the configuration on this strip. The difference between the 
energy of this long strip with periodic boundary conditions and the 
energy of the original configuration
is $O(U^{-9}) \nbad$. Each bad diamond costs an energy $O(U^{-7})$, 
so the energy of the original configuration is at least as large 
as that of our long strip. Finally,
in the single long strip we can apply Lemberger's argument to conclude
that the best arrangement is the most homogeneous one. 

\medskip

\noindent{\bf Proof of Theorem \ref{pthrm}:} 
We denote an ion configuration by $S$. We let $H(S)$ be the minimum 
energy of $H$ with this ion configuration, i.e., we minimize the 
Hamiltonian with respect to the electronic wave function.  If $U$ 
is sufficiently large, then the function $H(S)$ may be expanded in 
powers of $U^{-1}$. 
\begin{equation}
  H(S)= \sum_{m=2}^\infty U^{1-m} h_m(S)
 	\label{perturbseries}
\end{equation}
(Details of this perturbation theory may be found in \cite{gjlii,lem}.)
Because we are on a square lattice, 
only the even values of $m$ have a nonzero $h_m$. $h_m(S)$ is a local 
function. It is a sum of terms, each of whose support is contained 
in the support of a closed nearest neighbor walk with $m$ steps. 
We will let $H_m(S)$ denote $U^{1-m} h_m(S)$.

Let $M$ be the number of $3\times 3$ blocks in a 
configuration $S$ which are not a reflection or rotation of one of those 
shown in figure \ref{3blocks}. Kennedy \cite{k} showed that
\begin{equation}
	\alpha M U^{-3} \le H_2(S)+H_4(S)+U^{-3}L^2(a+b\rho)\le \gamma M U^{-1}
 	\label{4order}
\end{equation}
where $\alpha>0$, $\gamma>0$, $a$, and $b$ are constants. 
So after adding the constant $U^{-3}L^2(a+b\rho)$ to the Hamiltonian, 
we see that every $3\times3$ block which is not shown in figure 
\ref{3blocks} contributes a positive amount of energy at fourth order 
(or lower), and the configurations in which every $3\times3$ block is 
one of those in figure \ref{3blocks} have zero energy through fourth order.
\ifodd \preprint 
{
\begin{figure}[thp]
        \includegraphics{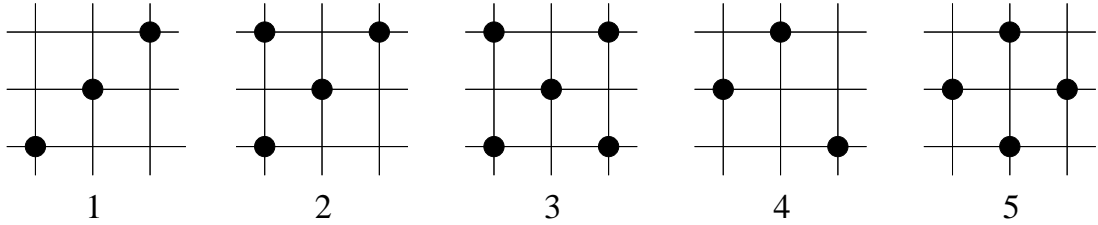}
        \caption{Configurations which minimize the energy locally.}
        \label{3blocks}
\end{figure}
}
\else {} \fi
The following lemma gives us a geometric interpretation of a configuration
in which every $3\times 3$ block is one of those shown in figure \ref{3blocks}.

\begin{lemma}{\bf (Watson \cite{watson})} 
Let S be a configuration in which every  
$3\times3$ block is one of those shown in figure \ref{3blocks}. Then there 
exists a tiling by squares and parallelograms as shown in figure
\ref{vtypes}, in which every vertex corresponds to an ion in $S$. 
\end{lemma}

\noindent{\bf Proof:} Let S be a configuration satisfying the hypothesis
of the lemma. Consider any occupied site. It is the center of a 
type 1, 2, or 3 block from figure  \ref{3blocks}.
Now consider a 3 by 3 block which shares
six sites with the original 3 by 3 block we considered. Since the configuration
at six of its sites is known and it must be one of the blocks in figure
\ref{3blocks}, there are only a few possibilities for the configuration
on the three unknown sites. By continuing in this manner and 
keeping track of all the cases one may 
determine the allowable local ion arrangements. These turn out to be rotations
and reflections of the three arrangements in figure \ref{vtypes}. $\qed$
\ifodd \preprint 
{
\begin{figure}[thp]
        \includegraphics{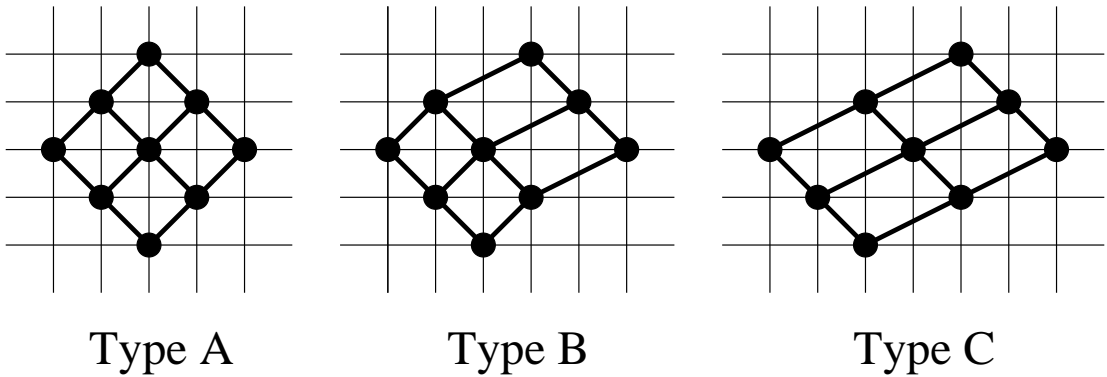}
        \caption{Possible local ion arrangements for a configuration in which
	every $3\times3$ block is one of those in figure \ref{3blocks}.}
        \label{vtypes}
\end{figure}
}
\else { } \fi

From the lemma and equation (\ref{4order}), 
we see that the configurations which 
minimize the energy through fourth order correspond to tilings by
squares and parallelograms. However, the densities one can obtain from
such tilings is restricted.
A configuration in which every ion is type C has density 1/3. One in which
every ion is type A has density 1/2. We may construct a configuration having
any rational density between 1/3 and 1/2 by placing an interface of slope
$\pm1$ between a
region of type A ions and a region of type C ions, the size of each region 
being chosen to give the desired density. The ions on the interface will
then be type B. So we see that from these square-parallelogram tilings,
one may obtain any rational density in $[1/3,1/2]$.
(Of course, there are many other ways to do so besides the construction 
we have given.)

Consider a square-diamond tiling and 
let $n_A,n_B,n_C$ be the number of type $A,B,C$ ions, respectively. 
These three integers do not determine the configuration, but they do
determine enough of the local structure of the configuration that 
the energy $H(S)$ may be computed through eighth order from 
$n_A,n_B$ and $n_C$. 
We show in the appendix that for $m=6$ and $8$  
\begin{equation}
	H_m(S)=U^{1-m}(n_A e^A_m + n_B e^B_m + n_C e^C_m)
        \label{hm}
\end{equation}
where 
\begin{equation}
e^A_6 = -3200,\qquad e^A_8 = 98000,\qquad e^B_6 = -1520,
\end{equation}
\begin{equation}
 e^B_8 = 33600,\qquad e^C_6=160,\qquad e^C_8=-15120.
\end{equation}
In a square diamond tiling, the numbers $n_A$, $n_B$, and $n_C$ are 
not independent. Since the area is $L^2$, we have
\begin{equation}
	2n_A + \frac{5}{2}n_B + 3n_C = L^2
	\label{parea}
\end{equation}
We are also keeping the number of ions fixed at $\rho L^2$, so
\begin{equation}
	n_A + n_B + n_C = \rho L^2.
	\label{pnumber}
\end{equation}
Using eqs. (\ref{parea}) and 
(\ref{pnumber}) to solve for $n_B$ and $n_C$ in terms of $n_A$, we have
\begin{displaymath}
	H_6(S)+H_8(S)-L^2f(U,\rho)=15680U^{-7}n_A.
\end{displaymath}
$f(U,\rho)$ is a function of $U$ and $\rho$, but there is no need to 
write it out explicitly,

The preceding paragraph assumed that $S$ was a tiling. 
In general, this will not be the case. 
Suppose $S$ is not a tiling and M is the number of $3\times3$ blocks
which are not rotations or reflections of those in figure \ref{3blocks}.
Then the number of sites in the region where $S$ is not a tiling is at
most $9M$. So the terms intersecting this region contribute at most 
$O(U^{-5})M$ to $H_6+H_8$. Equations (\ref{parea}) and (\ref{pnumber}) will no
longer hold, but the difference between the right and left hand sides will 
be bounded by a constant times $M$. Thus for a general $S$, we have
\begin{equation}
	H_6(S)+H_8(S)-L^2f(U,\rho)=15680U^{-7}n_A + O(U^{-5})M.
	\label{ph8}
\end{equation} 

We may now determine the ground states through order 8. From inequality 
(\ref{4order}) we see that any configuration with $M=0$ will minimize 
the energy through fourth order. In equation (\ref{ph8}), 
$M$ appears at order 6. 
So for large enough $U$ we want $M=0$. Since the coefficient of $n_A$ is 
positive in equation (\ref{ph8}), we also want $n_A=0$. So the 
configurations which minimize the energy through order 8 are those where
every ion is type B or C. These configurations correspond to 
square-parallelogram tilings with no adjacent $\pm 45^{\circ}$ strips of 
squares. They have zero energy through order 8 after we have subtracted
$L^2 f(U,\rho)$. 
All we have shown so far is that zero is a lower bound for the energy
through eighth order for densities in $[1/3,2/5]$. We can construct 
a configuration that attains this lower bound as follows. 
A region with only type B vertices has density $2/5$, while a region
with only type C has density $1/3$. By adjoining a type B region with 
a type C region with an interface between them at $45^{\circ}$ we can 
obtain any density in $[2/5,1/3]$ and attain our lower bound. 

If we just consider the energy through eighth order, then 
any configugation that has a ``bad'' $3\times 3$ block or a type A ion
will have energy higher than the ground state energy. 
The crucial point is that there exists a positive constant
$c$ so that difference of the energy and the ground state energy 
is at least $cU^{-7}$ 
times the number of these ``local defects'' in the configuration.
We now move on to the higher order terms.

Recall that $\rho = p/q \in [1/3,2/5]$ where $p$ and $q$ are relatively prime,
and the lattice $\Lambda$ is $L\times L$ where $L$ is a multiple of 
$4q$. We partition $\Lambda$ into diamonds containing 8 sites as 
shown in figure \ref{blocks}.
\ifodd \preprint 
{
\begin{figure}[thp]
        \includegraphics{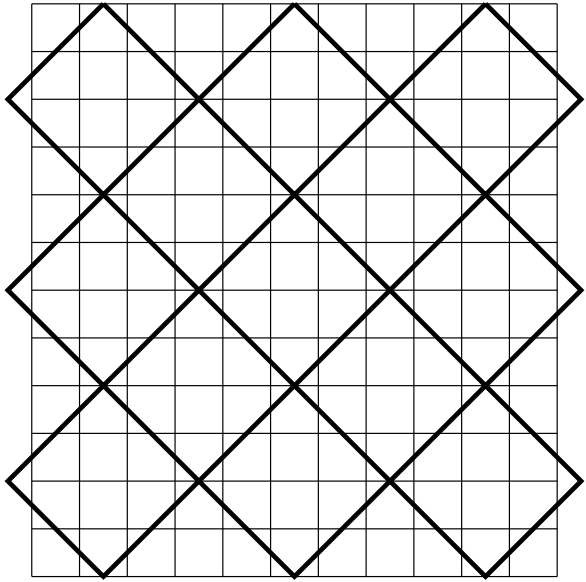}
        \caption{Partitioning of the lattice into diamonds containing 8 sites.}
        \label{blocks}
\end{figure}
}
\else { } \fi

We define a diamond $b$ to be $bad$ if it contains a type A ion or
intersects a bad $3\times3$ block. Otherwise, $b$ is $good$.  We let
$\nbad$ denote the number of bad diamonds. A $good$
diamond  is one which is contained in a region of the configuration
corresponding to a square-parallelogram tiling with no type A ions.
We know from the above that the energy through eighth order is bounded below
by a term proportional to the  number of bad  $3\times3$ blocks plus the 
number of type A ions, which is proportional to $\nbad$. So 
inequality (\ref{4order}) and equation (\ref{ph8}) tell us that
\begin{equation}
	H_2 + H_4 + H_6 + H_8 +U^{-3}L^2(a+b\rho)-L^2f(U,\rho)
	\ge c \nbad U^{-7}
	\label{ineq}
\end{equation}
where $c > 0$ is a constant. For future convenience, we define 
$$\tilde{H}=H+U^{-3}L^2(a+b\rho)-L^2f(U,\rho)
$$

Let $S$ be a configuration with density $\rho(S)\in[\frac{1}{3},\frac{2}{5}]$.
We label a parallelogram in $S$ with a $+$ if its short sides have slope $+1$, 
and a $-$ if its short sides have slope $-1$. Now, every good diamond must
contain a type B or C ion, both of which are the vertex of a parallelogram.
Any other parallelogram intersecting this diamond must have the same sign.
Thus the good diamonds may be labelled $+$ or $-$ in a natural way.
Furthermore, since all the ions in a good diamond are type B or C, two 
diamonds that share an edge must have the same sign. 
One may think of $S$ as being divided into regions where the
diamonds are $+$, regions where the diamonds are $-$, and regions of bad
diamonds. A $+$ region and a $-$ region must
be separated by a region of bad diamonds. 
Inside every $+$ region, $S$ is constant on lines of slope $+1$, and 
inside every $-$ region it is constant on lines of slope $-1$. 
We call this the $stripe$ property. 
More, however, may be said about
the arrangement of the occupied stripes inside each good region. Since 
every ion must be type B or C, each pair of occupied stripes must be 
separated by one or two unoccupied stripes. Also, since there are no type
A ions, we may not have consecutive stripes corresponding to the following 
occupation sequence: occupied, unoccupied, occupied, unoccupied, occupied. 
This rules out the checkerboard configuration, the sign of which is ambiguous. 
When a region has the stripe property with no type A ions, we say that the
stripes are $correctly$ $spaced$.  

We now divide up the $-$ regions into $45^{\circ}$ strips as follows. 
Choose any good diamond whose sign is $-$. As shown in figure \ref{extend},
we extend this set in the directions $\pm (1,1)$ until we hit a bad 
diamond. Every diamond in this $45^{\circ}$ strip is good, and, since one of
them is $-$, they must all be $-$. We may then carry our labelling one
step further and label this strip with a $-$. 
Note that every good $-$ diamond is contained in exactly one
such $-$ strip.  
\ifodd \preprint 
{
\begin{figure}[thp]
        \includegraphics{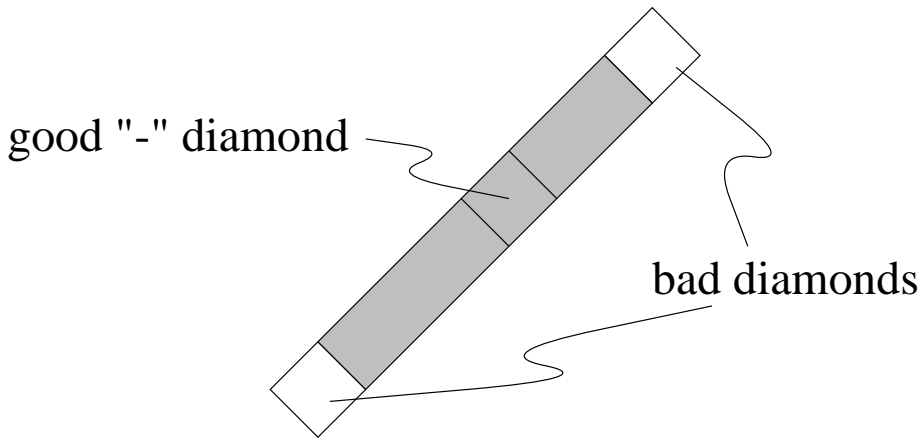}
        \caption{A good ``$-$'' diamond is chosen. The sites in the shaded
	region are contained in good ``$-$'' diamonds.}
        \label{extend}
\end{figure}
}
\else { } \fi
We divide up the $+$ region into $-45^{\circ}$ strips 
by changing all the signs in this construction.
We label these strips with a $+$. See figure \ref{rowcol}.
\ifodd \preprint 
{
\begin{figure}[thp]
        \includegraphics{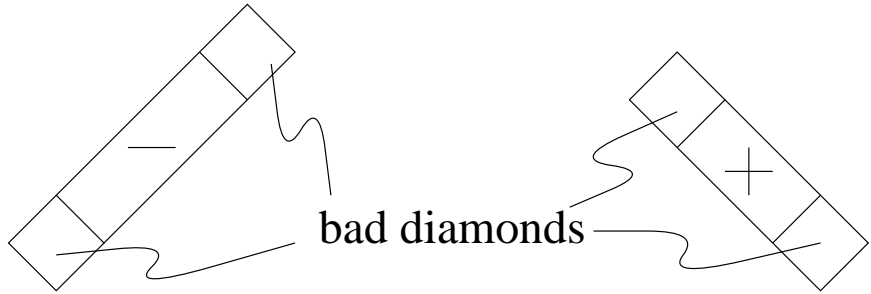}
        \caption{Left: a $-$ strip. Right: a $+$ strip. The 
		configuration inside the $-$ strip ($+$ strip) is constant 
		along lines of slope $-1$ ($+1$).}
        \label{rowcol}
\end{figure}
}
\else { } \fi

We now take the strips and rotate them so they are all horizontal and 
then glue them together to form one long strip.
Now rotate the bad diamonds and glue them onto one end to form an
even longer strip. 
This strip will contain all the sites of the original $\Lambda$.
This is shown in figure \ref{lstrip}. 
The strips have the stripe property. After rotation this means 
that in figure \ref{lstrip} the portion of the configuration 
that came from the strips is constant along vertical lines of lattice sites. 
In figure \ref{lstrip}, vertical lines of lattice sites only 
contain two sites, so this simply means that either both of these 
sites are occupied or both are empty.
In the portion of figure \ref{lstrip} 
which contains the bad diamonds, the configuration may not satisfy this 
property. However, in the statement of the theorem $L$ was assumed
to be a multiple of $4q$. Thus the total number of ions is a multiple of
$4p$. There are an even number of ions in each $+$ and $-$ strip, so there
must be an even number of ions in the bad region of our thin strip.
Therefore we may rearrange the configuration in this portion of the 
strip so it also has either zero or two ions in each vertical line
of lattice sites. We now glue the short
sides together and the long sides together to form a torus, call it $T$. 
Let $S_1$ denote the resulting configuration on $T$. We want to bound
$|\tilde{H}(S)-\tilde{H}(S_1)|$. As we shall see, we will only need to
estimate this difference at orders $\ge 10$. 
\ifodd \preprint 
{
\begin{figure}[thp]
        \includegraphics{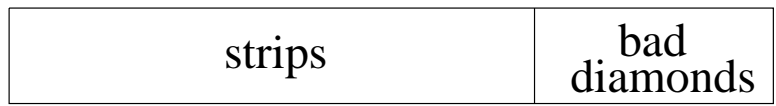}
        \caption{The thin strip obtained by gluing together the $+$ and $-$
	strips as well as the bad diamonds.}
        \label{lstrip}
\end{figure}
}
\else { } \fi

%%%%%%%%%%%%%%%%%%%%%%%%%%%%%%%%%%%%%%%%%%%%%%%%%%%%%%%%%%%%%%%%%%%%%%
%%%%%%%%%%%%%%%%%%%%%%%%%%%%%%%%%%%%%%%%%%%%%%%%%%%%%%%%%%%%%%%%%%%%%%
%%%%%%%%%%%%%%%%%%%%%%%%%%%%%%%%%%%%%%%%%%%%%%%%%%%%%%%%%%%%%%%%%%%%%%
%%%%%%%%%%%%%%%%%%%%%%%%%%%%%%%%%%%%%%%%%%%%%%%%%%%%%%%%%%%%%%%%%%%%%%
%%%%%%%%%%%%%%%%%%%%%%%%%%%%%%%%%%%%%%%%%%%%%%%%%%%%%%%%%%%%%%%%%%%%%%
%%%%%%%%%%%%%%%%%%%%%%%%%%%%%%%%%%%%%%%%%%%%%%%%%%%%%%%%%%%%%%%%%%%%%%
%%%%%%%%%%%%%%%%%%%%%%%%%%%%%%%%%%%%%%%%%%%%%%%%%%%%%%%%%%%%%%%%%%%%%%
%%%%%%%%%%%%%%%%%%%%%%%%%%%%%%%%%%%%%%%%%%%%%%%%%%%%%%%%%%%%%%%%%%%%%%

The perturbation theory leads naturally to an expression for $H$ as a sum
over nearest neighbor closed walks $\gamma$. 
\begin{displaymath}
	H=\sum\limits_{\gamma}w(\gamma)
\end{displaymath}
Lemberger gave a nice formula for the contribution of a walk:
\begin{equation}
\label{walkweight}
	w(\gamma)=\frac{(-1)^{m(S,\gamma)}}{|\gamma|}(2U)^{1-|\gamma|}\,
	{|\gamma|-2 \choose m(S,\gamma)-1}
\end{equation}
where $|\gamma|$ is the number of steps in $\gamma$ and $m(S,\gamma)$ is 
the number of ions at sites in the walk $\gamma$. 
Obviously, $H_n$ is equal to the sum over walks with $n$ steps.

We want to set up a correspondence between the terms in the 
perturbation series of $\tilde{H}(S)$ and $\tilde{H}(S_1)$.
So we need to set up a correspondence between walks on the torus
$T$ and the original square $\Lambda$.
We think of $Z^2$ as the covering space of the torus $T$. Then a 
walk $\gamma_1$ on $T$ has a natural lift to a walk $\hat \gamma_1$ on 
$Z^2$. Each site in $T$ comes from a site in $\Lambda$. We let 
$\gamma$ be the translate of $\hat \gamma_1$ such that $\gamma(0)$ is 
the site in $\Lambda$ that corresponds to the site $\gamma_1(0)$ in $T$. 
The map $\gamma_1 \rightarrow \gamma$ takes walks on $T$ to walks on 
$\Lambda$. 

There is a problem with this correspondence. 
The lifted walk $\hat \gamma_1$ need not return to its starting point. 
In fact the lifted walk will return to 
its starting point if and only if the original walk $\gamma_1$ is homotopic 
to zero. So we redefine $\tilde{H}(S_1)$ and 
$\tilde{H}(S)$ to be the sum of all the 
terms in the perturbation series that come from walks homotopic 
to zero. If a walk on $\Lambda$ is not homotopic to zero, then it
has at least $L$ steps. ($L$ is the length of the square $\Lambda$.)
Thus this changes $\tilde{H}(S)$ by terms of order $U^{-L}$. 
So the total change in $\tilde{H}(S)$ is of order $L^2 U^{-L}$. 
This is tiny even compared to the energy cost of a single bad diamond.
The change in  $\tilde{H}(S_1)$ is not so small since the torus $T$ is
quite narrow in one direction. But we are free to define 
$\tilde{H}(S_1)$ any way we wish. It merely serves to organize 
our proof.

Now suppose that $\gamma_1 \rightarrow \gamma$ 
and $\gamma$ does not encounter a bad 
diamond in $S$. Then $\gamma$ lies entirely in either a region 
of $+$ diamonds or a region of $-$ diamonds. Consider the first case. 
So the strips have slope $-1$ and the configuration is constant 
along lines with slope $+1$ in the vicinity of the walk $\gamma$. 
It follows that $\gamma$ sees the same configuration in $S$ that 
$\gamma_1$ sees in $S_1$, 
and so $\omega(\gamma)=\omega(\gamma_1)$.
Thus the terms in $\tilde{H}_n(S)-\tilde{H}_n(S_1)$ that do not 
cancel come from walks $\gamma$ in $S$ that do not encounter a bad diamond. 
So the number of terms that do not cancel is of order $n^2 N$. 
Hence we have 
\begin{equation}
\label{walkineq}
	\bigl|\tilde{H}_n(S)-\tilde{H}_n(S_1)\bigr|
	\le U^{1-n}c^n n^2\nbad
\end{equation} 

%%%%%%%%%%%%%%%%%%%%%%%%%%%%%%%%%%%%%%%%%%%%%%%%%%%%%%%%%%%%%%%%%%%%%%
%%%%%%%%%%%%%%%%%%%%%%%%%%%%%%%%%%%%%%%%%%%%%%%%%%%%%%%%%%%%%%%%%%%%%%
%%%%%%%%%%%%%%%%%%%%%%%%%%%%%%%%%%%%%%%%%%%%%%%%%%%%%%%%%%%%%%%%%%%%%%
%%%%%%%%%%%%%%%%%%%%%%%%%%%%%%%%%%%%%%%%%%%%%%%%%%%%%%%%%%%%%%%%%%%%%%
%%%%%%%%%%%%%%%%%%%%%%%%%%%%%%%%%%%%%%%%%%%%%%%%%%%%%%%%%%%%%%%%%%%%%%
%%%%%%%%%%%%%%%%%%%%%%%%%%%%%%%%%%%%%%%%%%%%%%%%%%%%%%%%%%%%%%%%%%%%%%
%%%%%%%%%%%%%%%%%%%%%%%%%%%%%%%%%%%%%%%%%%%%%%%%%%%%%%%%%%%%%%%%%%%%%%

% stuff moved to end of file 

%%%%%%%%%%%%%%%%%%%%%%%%%%%%%%%%%%%%%%%%%%%%%%%%%%%%%%%%%%%%%%%%%%%%%%

Note that although $S_1$ satisfies the stripe property, its stripes may not
be correctly spaced. The idea now is to form a finite sequence of 
configurations, $\{S_i\}_{i=1}^K$, such that 
\begin{displaymath}
	\big|\tilde{H}_n(S_{i+1})-\tilde{H}_n(S_i)\big| 
	\le U^{1-n}c^n, 
\end{displaymath} 
$K=O(\nbad)$, and the stripes
in the final configuration $S_K$ are correctly spaced. 
The strategy is as follows.
The reason why the stripes in $S_1$ may not be correctly spaced is 
because of the introduction of the bad diamonds, as well as the cuts. 
There are $O(\nbad)$ stripes in the bad region of $S_1$ and $O(\nbad)$ cuts,
so we may hope to correct the spacings with 
$O(\nbad)$ modifications, each modification costing 
at most $U^{1-n}c^n$ at order $n$. 

In making these modifications we must be sure that
each one fixes a ``fault'' in $S_i$ without creating a similar fault. All 
the modifications follow the same basic procedure. We illustrate this by 
considering a pair of adjacent occupied stripes in $S_1$. Since
the density of $S_1$ is $\le$ 2/5, if there is a pair of adjacent occupied 
stripes, there must be a pair of adjacent unoccupied stripes.
As shown in figure \ref{stwo}, 
we introduce two ``markers,'' one between the pair of adjacent occupied 
stipes and one between the pair of adjacent unoccupied stripes.
We let $W$ be the portion of the configuration between the markers. 
With the two axes of rotation $a$ and $b$ as shown in figure \ref{stwo},
we form $W^{-1}$ by rotating $W$ about the axis $a$. We then replace $W$
by $W^{-1}$ and, if necessary, rotate $W^{-1}$ about the axis $b$ so it fits
into place. (Whether or not we need to rotate about $b$ depends on 
the number of sites in $W$.)
$S_2$ is then the modified 
configuration. Any walk which gives a non zero 
contribution to $\tilde{H}_n(S_2)-\tilde{H}_n(S_1)$ must cross 
one of the markers, 
and so must visit one of the 8 sites adjacent to the markers. 
So we have that $\big|\tilde{H}_n(S_2)-\tilde{H}_n(S_1)\big| 
\le U^{1-n}c^n$.
\ifodd \preprint 
{
\begin{figure}[thp]
        \includegraphics{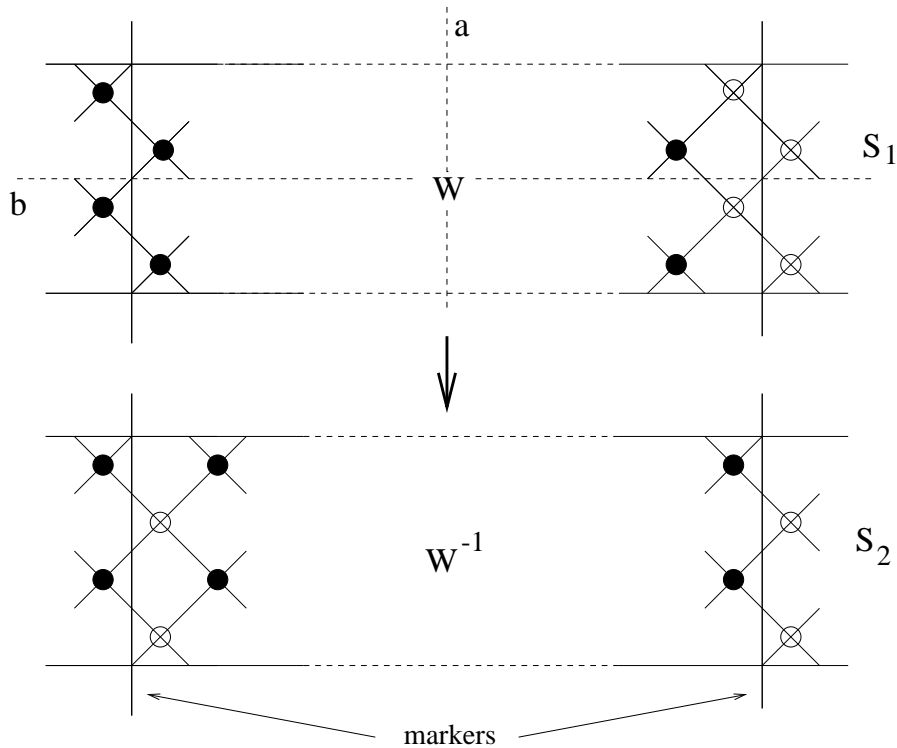}
        \caption{Separating a pair of neighboring occupied stripes.
          $W$ is rotated about axis $a$ and, if necessary,axis $b$.}
        \label{stwo}
\end{figure}
}
\else { } \fi
Our modification has allowed us to separate a neighboring pair of
occupied stripes without creating a new pair.
So we see that we can get rid of all such pairs with $O(\nbad)$ 
modifications. 

The next problem we focus on is the possibility of more than two 
consecutive unoccupied stripes. If this occurs, since the density is 
$\ge$ 1/3 there must be at least one pair of occupied stripes separated by
a single unoccupied stripe. Figure \ref{vacancy} shows how
we may reduce the number of adjacent unoccupied stripes
by one by a procedure similar to the above. 

\ifodd \preprint 
{
\begin{figure}[thp]
        \includegraphics[scale=.90]{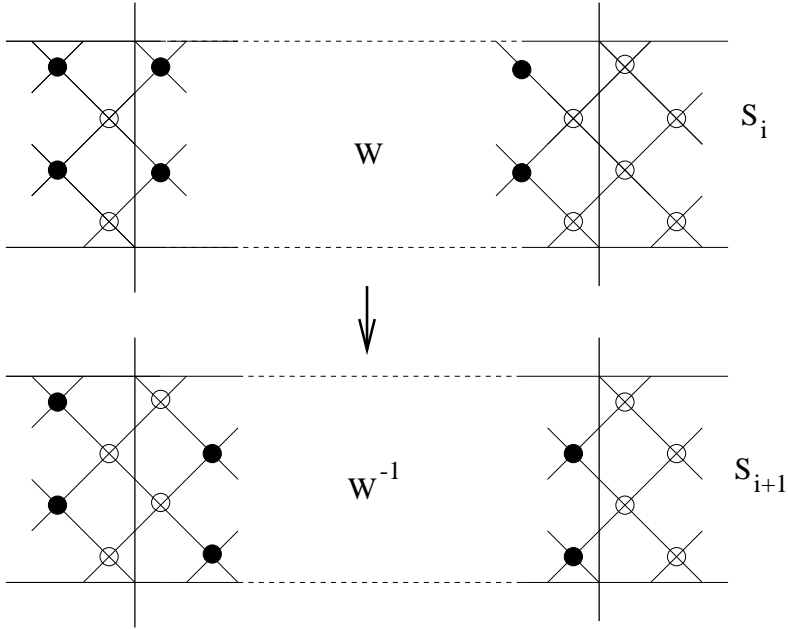}
        \caption{Reducing the number of consecutive unoccupied stripes.}
        \label{vacancy}
\end{figure}
}
\else { } \fi

After $O(\nbad)$ repetitions of the above procedures we reach a 
configuration in which every pair
of occupied stripes is separated by one or two unoccupied stripes. We now
face the possibility that our configuration contains type A ions. These ions
could have come from $S_1$ or they could have been formed by the 
modifications above. But we only made $O(\nbad)$ modifications, and $S_1$
contained at most $O(\nbad)$ type A ions. So our configuration can contain at
most $O(\nbad)$ type A ions. We let a ``1'' denote an occupied stripe and
a ``0'' denote an unoccupied stripe. The stripe sequence signifying a type 
A ion is then: 1, 0, 1, 0, 1. If this occurs, then since the density 
is $\le$ 2/5
there must exist the following stripe sequence: 1, 0, 0, 1, 0, 0, 1. In figure 
\ref{badass} we see how the procedure used above can get rid of these 
type A ions. 
At first glance, it may seem that this last modification, although removing
a type A ion at the left end of $W$ in $S_i$, may have created a type A ion
at the right end of $W^{-1}$ in $S_{i+1}$. However, if we do get a type
A ion in $S_{i+1}$ after the inversion of $W$, it must have been there before
the inversion, in $S_i$. Thus the number of type A ions is indeed reduced. 

\ifodd \preprint 
{
\begin{figure}[thp]
        \includegraphics[scale=.87]{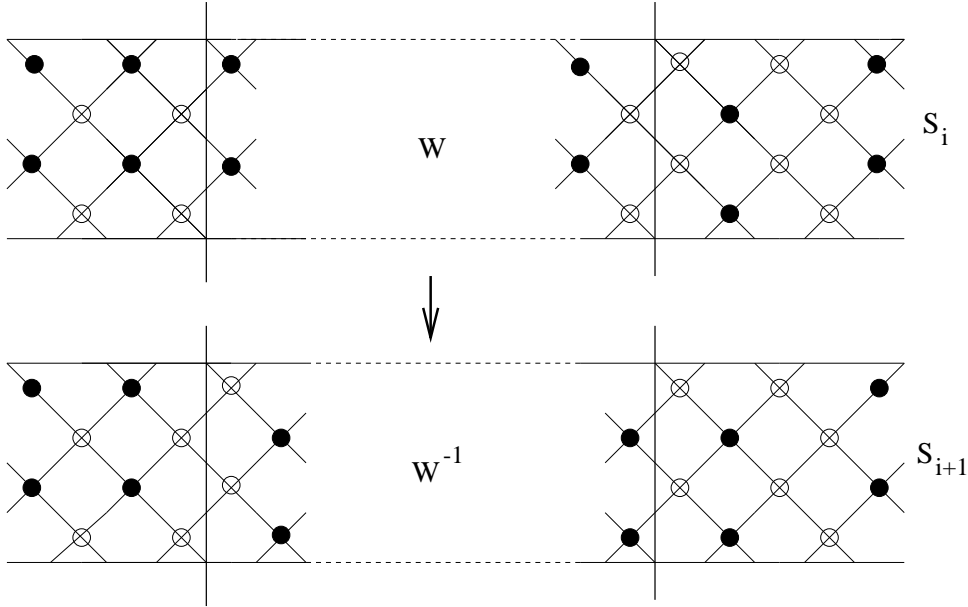}
        \caption{Removing the type A ions.}
        \label{badass}
\end{figure}
}
\else { } \fi

The last configuration in this sequence of modifications, $S_K$, has 
correctly spaced stripes. Furthermore, $K=O(\nbad)$ and 
$\big|\tilde{H}_n(S_{i+1})-\tilde{H}_n(S_i)\big| \le U^{1-n} c^n$ for 
each consecutive pair in our sequence. This gives us 
the following telescoping sum
\begin{equation}
\label{chuck}
	\tilde{H}(S)-\tilde{H}(S_K)=
	\bigl[\tilde{H}(S)-\tilde{H}(S_1)\bigr]+
	\sum\limits_{i=1}^{K-1} 
	\bigl[\tilde{H}(S_i)-\tilde{H}(S_{i+1})\bigr].
\end{equation}
The construction of our sequence above and inequality 
(\ref{walkineq}) imply that 
\begin{equation}
\label{blobby}
	\sum\limits_{even\ n\ \ge 10}\big|\tilde{H}_n(S) - 
	\tilde{H}_n(S_K)\big| 
	\le K \sum\limits_{even\ n\ \ge 10}U^{1-n}c^nn^2
	\le c\nbad U^{-9}
\end{equation}
In the last inequality we used the fact that $K$ is of order $N$. 
Since $S_K$ satisfies the stripe property and its stripes are correctly
spaced, we know that 
$$\sum\limits_{even\ n=2}^8\tilde{H}_n(S_K)=0$$ 
Equation (\ref{chuck}) and the bounds 
(\ref{ineq}) and (\ref{blobby}) imply
\begin{equation}
\label{ineq1}
	\tilde{H}(S) \ge \tilde{H}(S_K)+\alpha \nbad U^{-7}
	-c\nbad U^{-9}.
\end{equation}

Consider the torus $T$ on which $S_K$ is defined. Let $S_L$ be a 
configuration on $T$ which satisfies the stripe property and such that
the restriction of the configuration to a line in one of the lattice 
directions is the one-dimensional ground state for density $\rho(S)$ 
when $U$ is large as determined by Lemberger. ($S_L$ is unique up to lattice
symmetries.) We claim that among all the configuration on $T$ having 
density $\rho(S)$ and satisfying the stripe property, the energy is minimized
by $S_L$. In particular, the claim implies $H(S_K) \ge H(S_L)$. 

To establish the claim we show that there is a close relationship between 
the perturbation series for the two-dimensional model and that of 
the one-dimensional model. The terms in the Hamiltonian at order $2m$ 
may be written in terms of the walks as 
\begin{equation}
  H_{2m} = \sum_{\gamma: |\gamma|=2m} w(\gamma)
\end{equation}
with $w(\gamma)$ given by eq. (\ref{walkweight}). 
For the one-dimensional model the only change that needs to be made 
in this formula is that 
$\gamma$ is only summed over walks in the one-dimensional 
lattice. We will show that if the configuration has the stripe property 
then 
\begin{equation}
\label{twotoone}
  H_{2m} = c(m) {\sum_{\gamma: |\gamma|=2m, \gamma \, hor}} w(\gamma)
\end{equation}
where $\gamma \, hor$ means that the sum is over walks which
only take steps in the horizontal direction. So these walks are one-dimensional
although they may start at any point in the two-dimensional lattice. 
The combinatorial factor $c(m)$ is given by 
\begin{equation}
c(m)= \sum_{k=1}^m {m \choose k}^2
\end{equation}

To show (\ref{twotoone})
we think of a walk $\omega$ as a sequence of $2m$ letters 
chosen from L (left), R (right), U (up) and D (down). 
The number of L's must equal the number of R's, and the number of U's 
must equal the number of D's. 
(This is because of our redefinition of the Hamiltonian so that the 
perturbation series only contains terms coming from walks that are 
homotopic to zero.) 
Let $P \omega$ be the projection of the walk that is obtained by replacing
each U by an R and each D by an L. (This assumes that we are in the 
case where the slopes are $-1$. For slope $+1$, we replace U by L 
and D by R.) Then for configurations that satisfy 
the stripe property,  $H(\omega)=H(P \omega)$. 
Now let $\omega^\prime$ be a walk with only L and R steps. Since the walk 
is closed there are $m$ R's and $m$ L's. We obtain all walks $\omega$ with 
$P \omega= \omega^\prime$ by replacing some number of L's with D's
and the same number of R's with U's. If $k$ is the number of L's that we 
replace, then there are ${m \choose k}$ ways to make the L $\rightarrow$ D
replacements and an equal number of ways to make the R $\rightarrow$ U
replacements. So the number of $\omega$  with  $P \omega=\omega^\prime$ 
is $c(m)$ as defined above. 
Equation (\ref{twotoone}) says that for stripe configurations the 
two-dimensional Hamiltonian is the same as the one-dimensional Hamiltonian
order by order except for a multiplicative factor at each order. 
It is straightforward to check that Lemberger's argument applies 
unchanged when the $c(m)$ factors are included. 

The stripe arrangement in $S_L$ follows the Farey tree, just as the
ion arrangement in the 1-d case. So $S_L$ consists of a unit cell of
length $q$, which is periodically extended to cover the torus. Our 
original square lattice $\Lambda$ has sides whose lengths are multiples of 
$4q$. So we may build a configuration $S_L^\Lambda$ on $\Lambda$ which has
the same stripe arrangement as $S_L$, and satisfies 
$H(S_L^\Lambda)=H(S_L)$. We may then rewrite inequality (\ref{ineq1}) as
\begin{equation}
	H(S) \ge H(S_L^\Lambda)+\alpha\nbad U^{-7} -
	c\nbad U^{-9}.
	\label{sfinal}
\end{equation}
Assuming $U \ge \sqrt{c / \alpha}$, this shows that $S^\Lambda_L$ is a 
ground state. To see that these are the only ground states, we consider 
two cases. If $\nbad>0$ then the above bound says $H(S)>H(S^\Lambda_L)$.
If $\nbad=0$, then $S$ satisfies the stripe property with the stripes 
correctly spaced. So we know from Lemberger's argument that 
$H(S) \ge H(S_L^\Lambda)$, with equality holding only when 
$S=S_L^\Lambda$ up to a lattice symmetry. $\qed$

\newpage

\begin{appendix}

\section*{Appendix}

In this appendix we derive eq. (\ref{hm}) which expresses the sixth and 
eighth order terms in the Hamiltonian in terms of the number of 
ions of types A,B and C. 
Recall that the Hamiltonian at a given order is given in terms of 
walks by 
\begin{equation}
  H_{2m} = \sum_{\gamma: |\gamma|=2m} w(\gamma)
\end{equation}
This may be rewritten as 
\begin{equation}
  H_{2m} = \sum_{X} c_{m,X} S(X)
\end{equation}
where the sum is over subsets of the lattice and 
$S(X)$ is $1$ if the configuration $S$ has 
an ion at every site in $X$ and 0 otherwise.
 
Recall that $I(S)$ denotes the sets of sites at which there is an 
ion for the configuration $S$. 
If $S(X)$ is to give a non zero contribution to $H_6(S)+H_8(S)$, $X$ must
be a subset of $I(S)$ which is  
contained in a closed nearest neighbor walk of six or eight steps.
Let $S$ correspond to a tiling by squares and parallelograms.
Then the only
subsets $X\subset I(S)$ which can contribute at orders 6 and 8 are those 
shown, up to rotations and reflections, in figure \ref{bonds3}.
\ifodd \preprint
{ 
\begin{figure}[thp]
        \includegraphics[scale=.85]{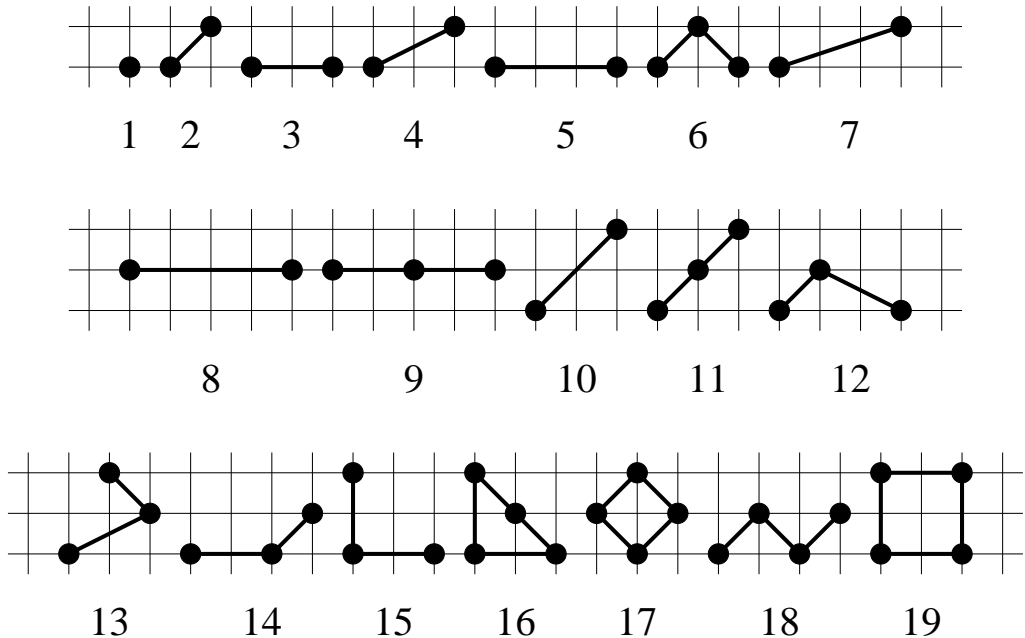}
        \caption{All possible subsets of $I(S)$ (up to lattice symmetries)
		which contribute to the energy at orders 6 and 8 when $S$
		corresponds to a square-parallelogram tiling.
		The lines serve to distinguish the subsets.}
        \label{bonds3}
\end{figure}
}
\else { } \fi
We will refer to an ion and the eight ions arranged about it as in 
figure \ref{vtypes} as the ``local arrangement'' of the ion. 
Recall that the local arrangement about an ion in $I(S)$ must look like one
of the three shown in figure \ref{vtypes}.
So for any $X$ which
gives a nonzero contribution to $H_6(S) + H_8(S)$, there is at least 
one ion $x$ such that $X$ is a subset of the local arrangement about $x$. 
There may be more than 
one such $x$. For example, if $X$ consists of a single ion, $X$ will be 
contained in 9 local arrangements. To compensate for this over counting, 
we define $\tilde{c}_{m,X} = \frac{c_{m,X}}{\kappa(X)}$ where $\kappa(X)$ is 
the number of ions whose local arrangement contains $X$. 
Now fix a type A ion and let $L$ denote the local arrangement about it.
Then for even $m \le 8$ we have  
\begin{displaymath}
	e^A_m=\sum\limits_{X \subseteq L} \tilde{c}_{m,X}
\end{displaymath}

This may be written as
\begin{displaymath}
        e^A_m=\sum\limits_{\stackrel{X\in figure\ \ref{bonds3}:}
	{X\subseteq L}}\tau_{A,X}\tilde{c}_{m,X}
\end{displaymath}
where ``$X\in figure\ \ref{bonds3}$'' means that $X$ is a rotation or 
reflection of one of the arrangements from figure \ref{bonds3}, 
and $\tau_{A,X}$
is the number of translates of $X$ appearing in the local configuration
about a type A ion. 
$e^B_m$ and $e^C_m$ are given by the same formula with $L$ taken to 
be the local arrangement of a type B or C ion respectively. 

Table \ref{table3} gives the coeffecients $c_{6,X}$ and $c_{8,X}$ 
for each $X$ in figure \ref{bonds3}.
The last three columns give the numbers 
$\frac{\tau_{A,X}}{\kappa(X)}$, 
$\frac{\tau_{B,X}}{\kappa(X)}$, and  
$\frac{\tau_{C,X}}{\kappa(X)}$.  
So we see, for example, that $e^A_6 = 64\times 1-384\times 2+96\times 2
-674\times 4=-3200$. 

We should note that one can use the above type of argument to 
show that an equation of the form (\ref{hm}) must hold. The values of 
$e^A_m, e^B_m$ and $e^C_m$ can then be computed by carrying out 
the perturbation theory for $H_m(S)$ for several different periodic 
configurations $S$ 
and then using the results to set up a linear system of equations 
for $e^A_m, e^B_m$ and $e^C_m$. We have done this as an independent check
on the calculation above.

\ifodd \preprint
{
\begin{table}[thp]
   \begin{center}
      \begin{tabular}{ || l || r | r || r | r | r ||}
         \hline
         $X$  & $c_{6,X}$  & $c_{8,X}$ 
         & $A$  & $B$ & $C$
        \\ \hline\hline
         1  & 64  & 112  & 1 & 1 & 1 \\ \hline
	 2  & -384  & 768  & 2 & 3/2 & 1 \\ \hline
         3  & 96  & -1360  & 2 & 1 & 0  \\ \hline
         4  & 216  & -768  & 0  & 1 & 2 \\ \hline
         5  & 24  & 448  & 0  & 1  & 2 \\ \hline
         6  & -672  & 11520  & 4  & 2  & 0  \\ \hline
	 7  & \  & 512  & 4 & 1 & 0  \\ \hline
         8  & \  & 32  & 2  & 0  & 0  \\ \hline
         9  & \  & 192  & 2  & 0  & 0  \\ \hline
         10  & \   & 1152  & 2  & 1 & 1 \\ \hline
         11  & \   & 768  & 2 & 1  & 1  \\ \hline
	 12  & \  & -1728  & 0  & 2  & 4 \\ \hline
         13  & \  & -5184  & 0  & 1 & 2 \\ \hline
         14  & \  & -384  & 8  & 2  & 0  \\ \hline
         15  & \  & -1248  & 4  & 1 & 0  \\ \hline
         16  & \  & 3840  & 4 & 1 & 0  \\ \hline
	 17  & \  & 23040  & 1 & 1/2 & 0  \\ \hline
         18  & \  & 3840  & 4  & 1 & 0  \\ \hline
         19  & \  & 960  & 1  & 0  & 0  \\ \hline
      \end{tabular}
      \caption{\protect Values of $c_{m,X}$ used to calculate the energy of 
		types A, B, and C ions at orders 6 and 8.}
      \label{table3}
   \end{center}
\end{table} 
}
\else { } \fi

\end{appendix}

\bigskip
\bigskip
\bigskip

\no {\bf Acknowledgements: } 
This work was supported by NSF grant DMS-9623509. 
It was part of Karl Haller's Ph.D. dissertation \cite{dissert}.

\ifodd \preprint 
{
\newpage}
\else 
{ 
\newpage
} 
\fi


\begin{thebibliography}{99}

\newcommand \jtype{\it}

\def \jsp    {{\jtype J. Stat. Phys.}\ }
\def \pr     {{\jtype Phys. Rev.}\ }
\def \prb    {{\jtype Phys. Rev. B}\ }
\def \prl    {{\jtype Phys. Rev. Lett.}\ }
\def \cmp    {{\jtype Commun. Math. Phys.} \ }
\def \jpc    {{\jtype J. Phys.: Condens. Matter}\ }
\def \jpa    {{\jtype J. Phys. A: Math. Gen.}\ }
\def \jpcold {{\jtype J. Phys. C: Solid State Phys.}\ }
\def \pl     {{\jtype Phys. Lett.}\ }
\def \lmp    {{\jtype Lett. Math. Phys.}\ }
\def \npb    {{\jtype Nucl. Phys. B}\ }
\def \jmp    {{\jtype J. Math. Phys.}\ }
\def \jap    {{\jtype J. Appl. Phys.}\ }
\def \jpsj   {{\jtype J. Phys. Soc. Jpn.}\ }
\def \rmp    {{\jtype Rev. Math. Phys.}\ }
\def \epl    {{\jtype Europhys. Lett.}\ }
\def \aihp   {{\jtype Ann. Inst. H. Poincar\'e}\ }
\def \jfa    {{\jtype J. Funct. Anal.}\ }
\def \zpcm   {{\jtype Z. Phys. B}\ }
\def \sc     {{\jtype Science}\ }

\newcommand \arctx{archived in Texas mp\_arc }
\newcommand \arclanl{archived in xxx.lanl }

\bibitem{bs} U. Brandt, R. Schmidt, Ground state properties of a 
spinless Falicov-Kimball model.
\zpcm {\bf 67}, 43 (1986).

\medskip

\bibitem{frob} N. Datta, R. Fern\'andez and J. Fr\"ohlich, Effective
Hamiltonians and phase diagrams for tight-binding models, 
\jsp {\bf 96}, 545 (1999).

\medskip

\bibitem{gjlii} C. Gruber, J. Jedrzejewski and P. Lemberger, 
Ground states of the spinless Falicov-Kimball model. II
\jsp {\bf 66}, 913 (1992).

\medskip

\bibitem{gm} C. Gruber and N. Macris, 
The Falicov-Kimball model: a review of exact results and extensions, 
Helv. Phys. Acta {\bf 69}, 851 (1996).

\medskip

\bibitem{triangular} C. Gruber, N. Macris, A. Messager and D. Ueltschi,
Ground states and flux configurations of the two dimensional Falicov-Kimball
model, \jsp {\bf 86}, 57 (1997).

\medskip

\bibitem{dissert} K. Haller,
Ground state properties of the neutral Falicov-Kimball model, 
Ph.D. dissertation, Program in Applied Mathematics, 
University of Arizona (1998). 

\medskip

\bibitem{k} T. Kennedy,
Some rigorous results on the ground states of the Falicov-Kimball model. 
{\jtype Rev. Math. Phys.} {\bf 6} 901-925 (1994).
Also in {\it The State of Matter}, Michael Aizenman and Huzihiro Araki
(eds.) World Scientific, 1994.

\medskip

\bibitem{ksep} T. Kennedy,
Phase separation in the neutral Falicov-Kimball model. 
\jsp {\bf 91}, 829-843 (1998).
\medskip

\bibitem{kl} T. Kennedy and E. H. Lieb, 
An itinerant electron model with crystalline or
magnetic long range order. {\jtype Physica A} {\bf 138}, 320-358 (1986).

\medskip

\bibitem{lem} P. Lemberger, Segregation in the Falicov-Kimball Model.
\jpa {\bf 25}, 715 (1992).

\medskip

\bibitem{mess} A. Messager, preprint. 

\medskip

\bibitem{mm} A. Messager and S. Miracle-Sol\'e, Low temperature states
in the Falicov-Kimball model, {\jtype Rev. Math. Phys.} {\bf 8}, 271 (1996). 

\medskip

\bibitem{miracle} S. Miracle-Sol\'e, A study of the large coupling
expansion in the Falicov-Kimball model, {\jtype Physica A } {\bf 232}, 
686-701 (1996).

\medskip

\bibitem{watson} 
G. I. Watson, Repulsive particles on a two-dimensional lattice,
{\jtype Physica A} {\bf 246}, 253 (1997).

\medskip

\bibitem{watlem} 
G. I. Watson and R. Lemanski, The ground-state phase diagram of the 
two-dimensional Falicov-Kimball model, {\jtype J. Phys. Condens. Matter} 
{\bf 7}, 9521 (1994).  

\end{thebibliography}
\end{document}